# ChemNODE: A Neural Ordinary Differential Equations Framework for Efficient Chemical Kinetic Solvers

Opeoluwa Owoyele, Pinaki Pal*

Energy Systems Division, Argonne National Laboratory, Lemont, IL 60439

*Contact author: pal@anl.gov, phone number: 630-252-1361

**Abstract**

Solving for detailed chemical kinetics remains one of the major bottlenecks for computational fluid dynamics simulations of reacting flows using a finite-rate-chemistry approach. This has motivated the use of fully connected artificial neural networks to predict stiff chemical source terms as functions of the thermochemical state of the combustion system. However, due to the nonlinearities and multi-scale nature of combustion, the predicted solution often diverges from the true solution when these deep learning models are coupled with a computational fluid dynamics solver. This is because these approaches minimize the error during training without guaranteeing successful integration with ordinary differential equation solvers. In the present work, a novel neural ordinary differential equations approach to modeling chemical kinetics, termed as ChemNODE, is developed. In this deep learning framework, the chemical source terms predicted by the neural networks are integrated during training, and by computing the required derivatives, the neural network weights are adjusted accordingly to minimize the difference between the predicted and ground-truth solution. A proof-of-concept study is performed with ChemNODE for homogeneous autoignition of hydrogen-air mixture over a range of composition and thermodynamic conditions. It is shown that ChemNODE accurately captures the correct physical behavior and reproduces the results obtained using the full chemical kinetic mechanism at a fraction of the computational cost.

## 1. INTRODUCTION

Chemical kinetic mechanisms for practical hydrocarbons fuels can contain as many as $O(10^2$-$10^3)$ chemical species and $O(10^3$-$10^4)$ reactions that describe the evolution of the species in time [1]. Computational fluid dynamics (CFD) simulations that employ such mechanisms and a finite-rate-chemistry combustion modeling approach require the integration of several partial differential equations (PDEs) to obtain the solutions of the reactive scalars, which evolve over a wide range of spatio-temporal scales. As a result, solving for detailed chemistry becomes the main bottleneck in CFD simulations of practical combustion systems that require millions of grid points [2]. Therefore, for practical engine-size geometries of interest to the automotive and aero-propulsion applications, highly simplified descriptions of chemistry are typically used to maintain computational tractability of simulations. However, these models sacrifice predictive accuracy. In line with these challenges, continued effort in developing efficient and scalable computational tools for capturing chemical kinetics is important. Such computational tools will lead to better model predictiveness at affordable computational costs, and ultimately aid the design and development of advanced energy systems.

Several studies have applied machine learning (ML) to solve these issues. Some of these studies address the *many species* problem by employing linear projections to derive a small set of reactive scalars from the original set of thermochemical scalars. In these studies, principal component analysis is typically used to project the original thermochemical space onto a low dimensional subspace where the system is defined by principal components. Regression-based ML models are then used to perform closure of chemical source terms, diffusion coefficients, and viscosity within the low-dimensional subspace. Past studies have involved *a priori* validation using data derived from experiments [3], direct numerical simulations (DNS) [4], one-dimensional turbulence (ODT) simulations [5], and homogenous reactors [6]. *A posteriori* validation studies have also been performed for an ODT simulation [7], DNS of premixed flames [8], and large eddy simulations (LES) [9] and Reynolds-averaged Navier Stokes (RANS) [10] simulations of the Sandia flames [11]. ML has also been applied to physics-based reduced-order models for turbulent combustion. In these studies, artificial neural networks (ANNs) have been used to learn flamelet tables involving two [12], three [13, 14], four [14, 15, 16], and five [14, 15, 16] control variables, and applied to problems such as the spray combustion [14, 17], combustion within a supersonic mixing layer [13], Sandia flame D [18], and piloted jet flames [19]. In another study, Cui et al.

[20] employed fully connected networks, optimized using a genetic algorithm, to predict ignition delay of n-butane/hydrogen mixtures in a rapid compression machine.

Other studies have used ML methods to accelerate the computation or integration of the chemical source terms. Christo et al. [21] used ANNs to represent a 4-step $H_2/CO_2$ chemical mechanism to perform simulations of a turbulent jet diffusion flame. Sen et al. [22] employed ANNs for modeling chemical kinetics by using them to predict the subgrid species source terms in the large-eddy simulation linear-eddy model (LES-LEM). In addition to these studies, neural networks have also been used to predict the chemical source terms of data-derived scalars within low-dimensional manifolds [7, 8, 23]. More recently, Ranade et al. [3] used neural networks to capture the process of pyrolysis of complex hydrocarbons, and Wan et al. [24] applied deep neural networks to the direct numerical simulation (DNS) of a syngas non-premixed oxy-flame. Ji and Deng [25] developed a physics-informed chemical reaction neural network (CRNN) for extracting expressions for reaction rates from data, while Barwey and Raman [26] developed a neural network-inspired formulation for computing chemical kinetics on graphics processing units (GPUs).

The overall methodology of source term predictions using ML methods in these previous studies is as follows. First, data is generated by running simulations with a complex chemical mechanism. By learning from samples obtained from these simulations, the ML model learns to predict the source terms as functions of the thermochemical state. If the errors between the predicted and actual source terms are below an acceptable threshold, it is assumed that the neural network can be coupled with a numerical solver and integrated to recover the true solution. Oftentimes, however, the predicted solution diverges from the true solution or becomes unstable when coupled with a numerical solver. Since combustion is a highly nonlinear phenomenon, even small errors in the predictions of the source terms, especially if occurring during an early time instance, can lead to very erroneous solutions.

In the present work, we proposed a novel approach toward ML-based estimation of chemical kinetics. As opposed to separating the data-driven learning and numerical validation phases, the approach used in the study combines both within an integrated framework. This approach, termed as ChemNODE, captures chemical kinetics using a recent class of deep learning models known as neural ordinary differential equations (NODEs) [27]. ChemNODE calculates the loss function based on the actual and predicted solutions and directly learns to predict the chemical source terms that lead to an accurate ordinary differential equation (ODE) solution. The following sections include a description of the ChemNODE approach, a proof-of-concept study for a canonical hydrogen-air homogeneous autoignition problem, and a discussion of the approach from an optimization perspective. The paper ends with a summary of the main conclusions and authors' outlook on directions for future work to further improve the training and performance of ChemNODE.

## 2. ChemNODE framework

The configuration considered in this work is an unsteady homogenous reactor with no diffusion or convection. For a system of $N$ thermochemical scalars, the vector containing the chemical source terms of the thermochemical scalars is denoted as $\dot{\boldsymbol{\omega}} = (\dot{\omega}_1, \dot{\omega}_2, \ldots, \dot{\omega}_N)^T \in \mathbb{R}^{N \times 1}$. The thermochemical state (comprised of species mass fractions and temperature) of the system at an arbitrary time instant is denoted as $\boldsymbol{\theta} = (\theta_1, \theta_2, \ldots \theta_N)^T \in \mathbb{R}^{N \times 1}$, while the column matrix containing all the observations for an arbitrary species (or temperature) is given as $\boldsymbol{\psi}$. Based on these, we can denote a matrix containing the $S$ observations of $N$ thermochemical scalars as $\boldsymbol{\Psi} = (\boldsymbol{\theta}_1, \boldsymbol{\theta}_2, \ldots, \boldsymbol{\theta}_s)^T = (\boldsymbol{\psi}_1, \boldsymbol{\psi}_2, \ldots, \boldsymbol{\psi}_N) \in \mathbb{R}^{S \times N}$. The evolution of the thermochemical scalars is then given by:

$$\frac{d\boldsymbol{\theta}}{dt} = \frac{\dot{\boldsymbol{\omega}}(\boldsymbol{\theta}; t)}{\rho} \tag{1}$$

The chemical source term, $\dot{\boldsymbol{\omega}}$, is a function of the thermochemical state of the system at any given time, $t$. The chemical source terms are typically obtained from chemical mechanisms that are formulated based on the law of mass action. While global formulations of chemical kinetics that assume a single reaction step between major reactants and products have been used in many previous studies to capture chemical kinetics, this is an extreme simplification of many practical combustion devices. In reality, many species participate in several elementary reactions, and these need to be computed to obtain the chemical source terms. While such detailed treatment of chemical kinetics is desirable, it leads to extreme computational costs. Therefore, in this work, the computation of source terms using chemical mechanisms is replaced with ANNs. Eq. 1, written in terms of an arbitrary species, $k$, becomes

$$\frac{dY_k}{dt} = \mathcal{M}_k(\boldsymbol{\theta}, \boldsymbol{\xi}_k) \quad \text{for } k = 1, 2, \ldots, N \tag{2}$$

$\mathcal{M}_k$ represents the output of a neural network with one hidden layer and a sigmoid activation (used in this work), $\sigma$, and is given by:

$$\mathcal{M}_k = \sigma(\mathbf{W}_{1,k}\boldsymbol{\theta} + \mathbf{b}_{1,k})\mathbf{W}_{2,k} + \mathbf{b}_{2,k} \tag{3}$$

Unique neural networks, each defined as described in Eq. 3, are used to predict the source terms of different thermochemical scalars. $\boldsymbol{\xi}_k = \left(\mho(\mathbf{W}_{1,k}), \mho(\mathbf{b}_{1,k}), \mho(\mathbf{W}_{2,k}), \mho(\mathbf{b}_{2,k})\right)$ is a vector that contains the trainable parameters of the neural network for the $k$th thermochemical scalar, with $\mho$ being an operator that vectorizes a matrix with $m$ rows and $n$ columns via the following mapping: $\mathbb{R}^{m \times n} \rightarrow \mathbb{R}^{mn}$. The $\mathbf{W}$'s and $\mathbf{b}$'s are the weights and biases of the neural network, which are initialized using the desired probability distribution and progressively tuned during training to obtain better predictions. In other words, the process of training a neural network is an optimization problem, where the goal is to find the weights and biases that minimize a loss function of interest. For regression-type problems, this loss function is typically a measure of the error between the predicted and actual values of the target variable.

The conventional practice when developing ML models for reacting flow simulations starts by generating data that covers a space of interest. While many earlier proof-of-concept studies performed training and validation studies using the same CFD configuration, more recent studies have generated data using approaches such as stochastic micro-mixers [24] and a variant of the pairwise mixing stirred reactor [23, 28]. Using these relatively inexpensive simulations, training samples are collected at various spatiotemporal locations to form a training database, consisting of various thermochemical states and their corresponding source terms. After performing some type of dimensionality reduction, either by combining or eliminating variables, a neural network is trained to learn the source terms as functions of the thermochemical state by using samples from the database. After training, the neural network library is coupled with a numerical solver to compute the source terms during unsteady CFD simulations. The problem is one of finding the weights and biases that minimize the following loss function:

$$\mathcal{L} = \left\|\boldsymbol{\Omega} - \widehat{\boldsymbol{\Omega}}\right\|_2^2 \tag{4}$$

In Eq. 4, $\boldsymbol{\Omega} \in \mathbb{R}^{S \times 1}$ contains $S$ observations of a source terms of an arbitrary thermochemical scalar, as obtained from the full chemical mechanism. The $\widehat{\cdot}$ symbol corresponds values predicted by the ML model, with an architecture as described in Eq. 3. On the other hand, the approach used in this study differs in that it involves finding the weights and biases that minimize:

$$\mathcal{L} = \|\boldsymbol{\psi} - \widehat{\boldsymbol{\psi}}\|_2^2 \tag{5}$$

where $\boldsymbol{\psi}$ contains the observations of an arbitrary thermochemical scalar, obtained at various time instants during the integration of the ODE in Eq. 1. On the other hand, $\widehat{\boldsymbol{\psi}}$ is the predicted solution, obtained by integrating Eq. 2. As such, the loss function in Eq. 5 measures the difference between the actual and predicted solutions. This is in contrast to Eq. 4 where the loss indicates how well the neural network predicts the chemical source terms. An illustration of the ChemNODE approach used in this study is shown in Fig. 1.

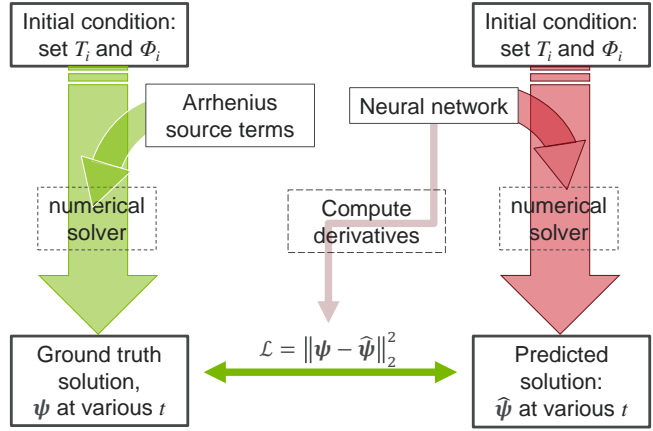

Figure 1. *Illustration of the machine learning approach used by ChemNODE.*

The ChemNODE approach is also described in Algorithm 1. In Algorithm 1, $\boldsymbol{x}$ represents the initial state used to train for each species. In the course of this study, it was found that attempting to learn the source terms for all species concurrently was too unstable because of the nonlinear dependencies that exist between the species. Instead, a progressive approach to training was employed, where the species were trained sequentially. At each value of *i,* some of the thermochemical scalars have trained neural networks, while others only have the ground truth solution. Therefore, for a given value of *i* in algorithm 1, $\boldsymbol{x}$ is a concatenation of two vectors, $(\hat{\theta}_1, \hat{\theta}_2, \ldots, \hat{\theta}_i)$ and $(\theta_{i+1}, \theta_{i+2}, \ldots, \theta_N)$. The ChemNODE approach used in this study was implemented in Julia Language [29].

---

**Algorithm 1** ChemNODE

---

**Input**: number of time instants: *S*, number of thermochemical scalars: *N*
**Input**: $\boldsymbol{\Psi} \in \mathbb{R}^{S \times N}$: ground-truth solution of thermochemical state at various time instants
**Require:** $I_S$: a numerical ODE solver that advances the solution in time
**for** $i = 1$ to $N$ >> loop over all thermochemical scalars
   Initialize $\boldsymbol{\xi}_i$ >> initialize the model parameters for the *i*th species

    **repeat**
        $x \leftarrow (\hat{\theta}_1, \hat{\theta}_2, \ldots, \hat{\theta}_i, \theta_{i+1}, \theta_{i+2}, \ldots, \theta_N)$ where $\boldsymbol{\theta}$ is evaluated at time $t = 0$    >> define the initial state
        $\widehat{\boldsymbol{\Psi}} \leftarrow I_S(\mathcal{M}(x, \xi_i))$    >> integrate the system to obtain the predicted solution
        $\boldsymbol{\varepsilon} \leftarrow (\mathbf{e}_1^T, \mathbf{e}_2^T, \ldots, \mathbf{e}_S^T)^T$ where $\mathbf{e}_j = (\boldsymbol{\theta}_j - \widehat{\boldsymbol{\theta}}_j); j = \{1, 2, \ldots, S\}$    >> calculate the error, concatenate, and store as a column vector
        $\lambda \leftarrow 0.1\sqrt{\|\boldsymbol{\varepsilon}\|_2}$    >> compute the damping factor
        $\mathbf{H} \leftarrow \mathbf{J}_{\boldsymbol{\varepsilon}}(\xi_i^t)$    >> compute the approximate Hessian
        $\xi_i^{t+1} \leftarrow \xi_i^t - (\mathbf{H} + \lambda \mathbb{I})^{-1} \mathbf{J}_{\boldsymbol{\varepsilon}}(\xi_i)\boldsymbol{\varepsilon}$    >> update weights
    **Until** $\xi_i^t$ is converged
    $\xi_i \leftarrow \xi_i^{t+1}$
  **end for**

The ChemNODE approach involves computing derivatives, not only through the neural network layers but also the operations of the ODE solver. To train the neural network by minimizing the loss function described in Eq. 5, the derivative of the ODE solution with respect to the neural network weights and biases needs to be computed. There are a number of ways in which this has been done in the literature. One approach involves a method known as adjoint sensitivity analysis [30], which involves framing an auxiliary ordinary differential equation whose solution gives the derivatives of the solution with respect to the neural network parameters. The `solution can be obtained by solving this auxiliary ODE backward in time [27], but this approach suffers from extreme integration errors under certain conditions. The ODE can also be solved by performing multiple forward passes [31, 32], a process that can be made more efficient by using a checkpointing scheme [33]. In this study, due to the small size of the neural networks, the sensitivity is calculated using a forward-mode continuous sensitivity analysis from Julia's DiffEqSensitivity.jl [34] package. To calculate gradients, a forward mode automatic differentiation [35] in Julia's ForwardDiff.jl [36] was employed. Julia's forward-mode automatic differentiation calculates derivatives by tracking atomic functions of the neural network and numerical solver, and propagating derivatives along with the computation of the solution. Numerical integration during ChemNODE training was performed using an A-L stable stiffly-accurate *4*th order ESDIRK method from Julia's DifferentialEquations.jl library [37].

### 3. Results

### 3.1 Homogenous hydrogen-air reactor

As a proof-of-concept study to test the capabilities of ChemNODE to accurately capture chemical kinetics, the problem of a homogenous zero-dimensional reactor at constant pressure is considered. Typically, in reacting flow CFD solvers, chemical kinetics is decoupled from the convective and diffusive effects during integration using a Strang splitting approach, wherein chemistry is solved in each computational cell assuming it to be a homogeneous reactor. Therefore, the canonical configuration chosen for this study – a homogeneous reactor with no transport effects – is relevant and compatible with 3D CFD solvers. Separate neural networks, each consisting of a single layer and 10 neurons, were trained for each species. In this study, the results are based on hydrogen-air autoignition at 1 atm. Using the formulation in Eq. 1, data was generated on a grid using various values of initial equivalence ratios, $\phi_i$ and initial temperatures, $T_i$ using Cantera [38], a chemical kinetics suite. During the data generation phase, numerical integration was performed using a variable-coefficient ODE (VODE) solver [39], which works by switching between the implicit Adams solver for non-stiff systems and a backward differentiation formula for stiff systems. The composition space used for training the neural network involved running all reactors to steady-state. The initial temperature, $T_i$, was varied between 950 $K$ and 1200 $K$, while the equivalence ratio, $\phi_i$, was varied from 0.5 to 1.5. The training dataset in this study consisted of 36 time-series generated using different initial temperatures and equivalent ratios. For each time series, the solution was saved at 50 points in time during integration. Before training, species with negligible concentrations, $H$, $HO_2$, and $H_2O_2$ were excluded from the thermochemical state for training. This led to a thermochemical vector consisting of temperature and 6 species: $\boldsymbol{\theta} = (T, H_2, O_2, OH, O, H_2O, N_2)^T$.

Figures 2– 4 show comparisons of the solutions obtained from ChemNODE and those obtained using the full chemistry mechanism. In the figures, the plots on top are for the temperature and reactants, which appear like progress variables even when plotted on a logarithmic scale. At the bottom, the radicals $O$ and $OH$, and the product of combustion, $H_2O$, are shown. It must be noted that the scalars in the top plots are normalized because temperature exists on a different scale from the species, while the plots at the bottom are raw values. In all the plots, the logarithmic values of the scalars are displayed. The lines represent the actual solution, while the symbols represent the predicted solution.

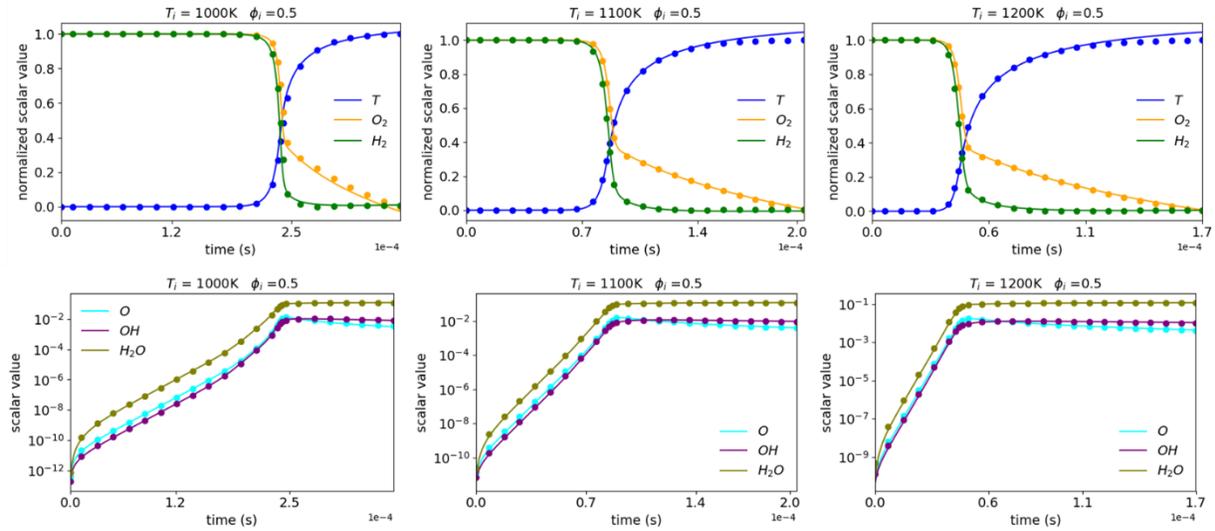

Figure 2. *Comparison of ChemNODE and detailed chemical mechanism solutions at $\Phi_i = 0.5$. The lines represent the actual solution, while the symbols represent the predicted solution.*

Figure 2 shows the evolution of the chemical species at a fuel-lean condition of $\Phi_i = 0.5$, for various initial temperatures, $T_i$. From the figures, it can be clearly seen that ChemNODE captures the correct behavior under different conditions. The trends for temperature and the reactants (top), as well as the other species (bottom) are well-captured. In general, there is a lag between the zone of rapid $O_2$ and $H_2$ consumption – the oxidizer lags behind the fuel slightly. ChemNODE accurately captures this physical behavior. There is an underprediction in the final temperature for all initial temperatures, but the maximum absolute relative error that occurs is only about 3.0%. Overall, the average mean absolute error (AMAE) normalized by the mean values of the species over the temperatures considered in the figure are $7.14 \times 10^{-4}$, $4.72 \times 10^{-3}$, $2.18 \times 10^{-3}$, $9.14 \times 10^{-3}$, $8.51 \times 10^{-3}$, and $1.17 \times 10^{-2}$ for $T$, $H_2$, $O_2$, $O$, $OH$, and $H_2O$, respectively.

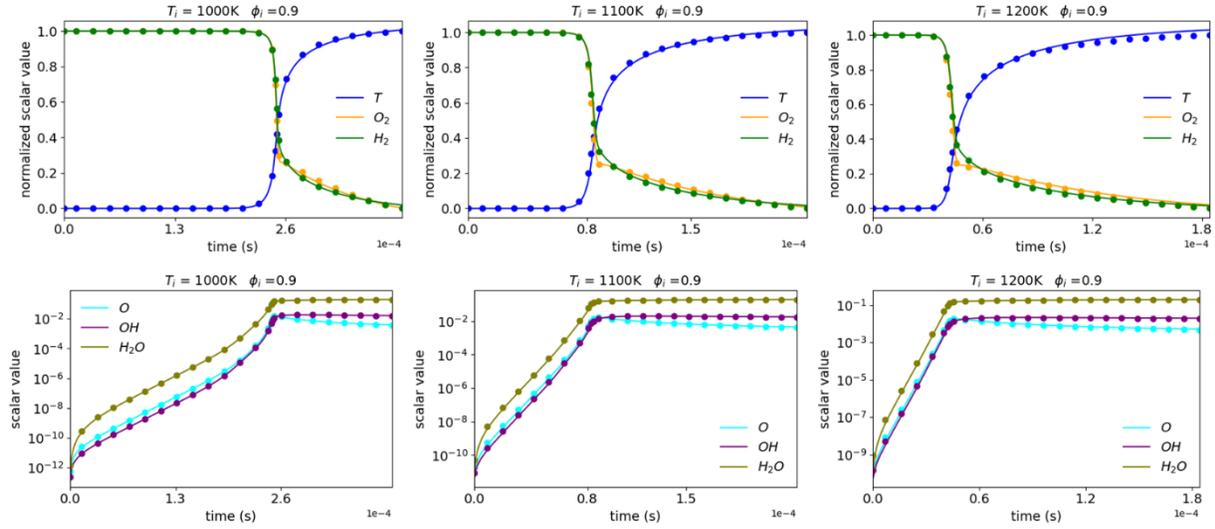

Figure 3. *Comparison of ChemNODE and chemical mechanism solution at $\Phi_i = 0.9$. The lines represent the actual solution, while the symbols represent the predicted solution.*

Figure 3 shows the evolution of the chemical species at a slightly fuel-lean condition of $\Phi_i = 0.9$, at various initial temperatures. Once more, it can be seen that ChemNODE captures the correct behavior. While the final temperature is once more underpredicted by 1.3% on average, the AMAE in the prediction of the species is $2.51\times10^{-3}$, $3.40\times10^{-3}$, $9.72\times10^{-3}$, $1.00\times10^{-2}$, and $1.44\times10^{-2}$ for $H_2$, $O_2$, $O$, $OH$, and $H_2O$, respectively. Figure 4 shows the same information as Figs. 2 and 3, but at a fuel-rich condition of $\Phi_i = 1.5$. Here, the consumption of the fuel, $H_2$, lags behind the consumption of the oxidizer. This phenomenon is well-captured by ChemNODE. Overall, the AMAE of the species over the temperatures considered in the figure are $8.70\times10^{-4}$, $2.76\times10^{-3}$, $7.34\times10^{-3}$, $9.20\times10^{-3}$, $9.96\times10^{-3}$, and $1.55\times10^{-2}$ for $T$, $H_2$, $O_2$, $O$, $OH$, and $H_2O$, respectively.

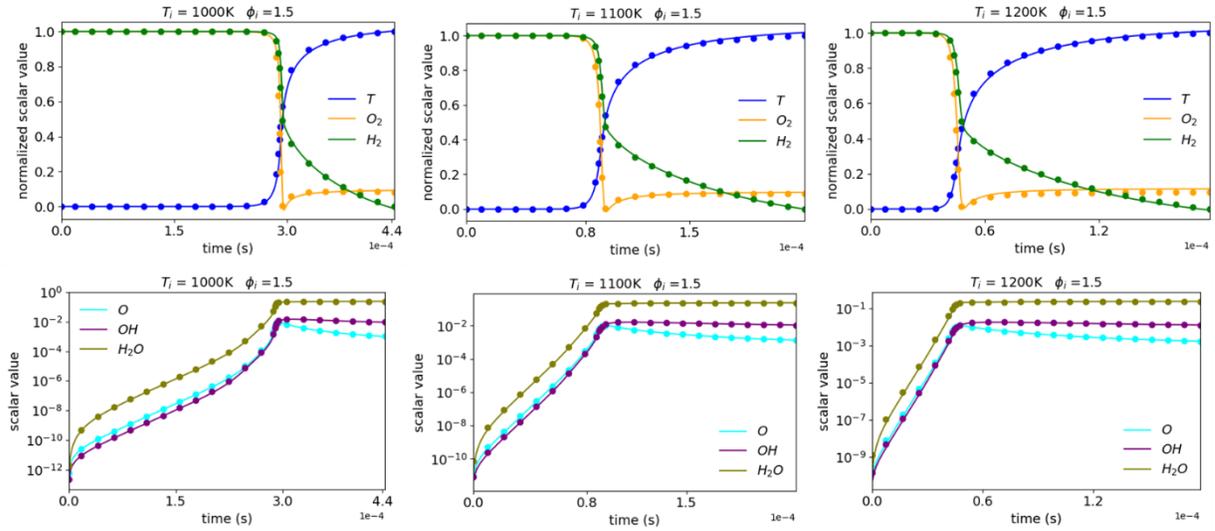

Figure 4. *Comparison of ChemNODE and chemical mechanism solution at $\Phi_i = 1.5$. The lines represent the actual solution, while the symbols represent the predicted solution.*

Figure 5 shows the ignition delay as a function of the initial equivalence ratio, $\Phi_i$, at different initial temperatures, $T_i$. The red symbols are the actual values while the blue circles are the predicted values. Here, the ignition delay is defined as the time when the maximum rate of temperature increase occurs during combustion. At higher values of $T_i$, the ignition delay profile has a *u*-shaped profile with respect to the equivalence ratio. At lower values of $T_i$, however, the ignition delay increases monotonically with equivalence ratio. From Fig. 5, it is evident that the ignition delay under various conditions is accurately predicted by ChemNODE. Overall, the results show that ChemNODE can capture both monotonically (e.g., $T$, $H_2$) and non-monotonically (e.g., $O$, $OH$) behaving reactive scalars.

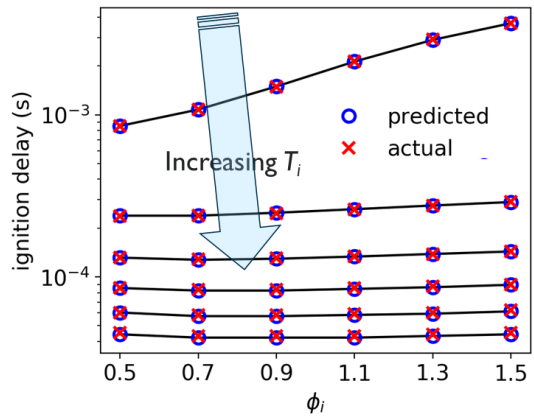

Figure 5. *Ignition delay of hydrogen-air mixture as a function of $T_i$ and $\Phi_i$.*

Finally, Fig. 6 shows a comparison of the total time taken to advance the solution to steady-state using ChemNODE and the full chemical mechanism with initial conditions selected across various values of $T_i$ and $\Phi_i$. Even for a small mechanism involving hydrogen-air combustion used in the present work, ChemNODE notably leads to a speedup of about 2.3x over the full mechanism. This would translate to the same speedup factor for ChemNODE relative to the full hydrogen-air chemical mechanism in a CFD simulation. It can be expected that for higher hydrocarbon fuels with much larger number of species and chemical reactions, more significant savings will be achieved with ChemNODE. This is because such mechanisms are highly complex and contain more redundancies that could be potentially eliminated in a data-driven reduction approach, such as is done in ChemNODE.

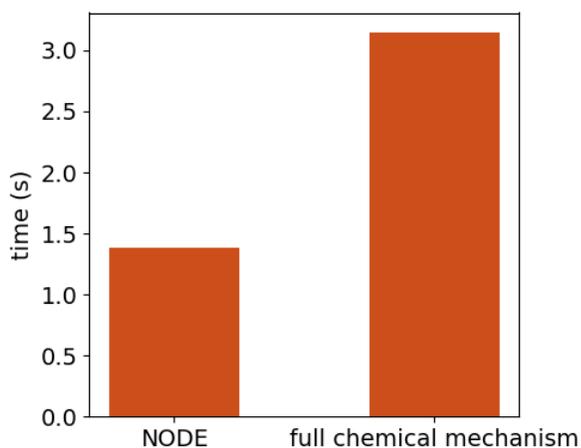

Figure 6. *Time taken to reach steady-state by ChemNODE and the detailed chemical mechanism.*

**3.2 Comparison between first-order and second-order optimization**

In this section, we examine the problem of interest from an optimization perspective and offer some suggestions on the types of optimization algorithms that are suitable for neural ODEs in chemical kinetics modeling. Optimization approaches for machine learning can broadly be classified into first-order and second-order approaches. First-order approaches assume that updates to the weights can be obtained via first-order derivatives of the loss with respect to the neural network parameters. While the most basic approach involves taking a step in the direction of steepest descent, in general, first-order optimizers can be thought of as updating the trainable parameters of the neural network according to the following:

$$\boldsymbol{\xi}_{new} = \boldsymbol{\xi}_{old} - f\big(\nabla_\theta \mathcal{L}(\boldsymbol{\xi}_{old})\big). \tag{6}$$

In Eq. 6, $f$ is a function that depends on the first-order optimization algorithm. For gradient descent, for instance, $f\left(\nabla_{\xi}\mathcal{L}(\xi_{old})\right) = \lambda \nabla_{\xi}\mathcal{L}(\xi_{old})$, where $\lambda$ is the learning rate. For problems with extreme non-convexity or ill-conditioned Hessians, first-order algorithms that take a fixed step in each direction can get stuck in local minima or take extremely long to converge. On the other hand, second-order optimizers approximate or compute the curvature of the optimization surface. In this study, the results shown in section 3.1 were obtained using a modified second-order Levenberg-Marquardt [40] (LM) optimizer, where the weights were updated according to:

$$\xi_{new} = \xi_{old} - (H + \lambda \mathbb{I})^{-1} J_e(\xi) e, \tag{7}$$

where $J_e(\xi) = \partial e / \partial \xi$, is the Jacobian matrix, $H = J^T J$ is the estimated Hessian, $e = y - \hat{y}$ is the difference between the actual and predicted solution, and $\lambda = 0.1\sqrt{e.e}$ is the damping factor.

To visualize the loss function surface, an approach similar to those introduced in previous studies [41, 42] is used here. The following steps are followed.

1. ChemNODE is trained using the desired optimizer until it reaches a steady-state solution, where the loss function stops decreasing. The best neural network parameters are denoted as $\xi_b$.
2. A random vector, $\eta$, that has the same number of entries as $\theta$ is chosen.
3. Compute the loss as we move along a straight line from the edge of $\eta$ to that of $\theta_b$, and onwards to the edge of the vector $2\xi - \eta$. Compute $\xi = \xi_b + \alpha\eta$, for $\alpha \in [-1, 1]$.

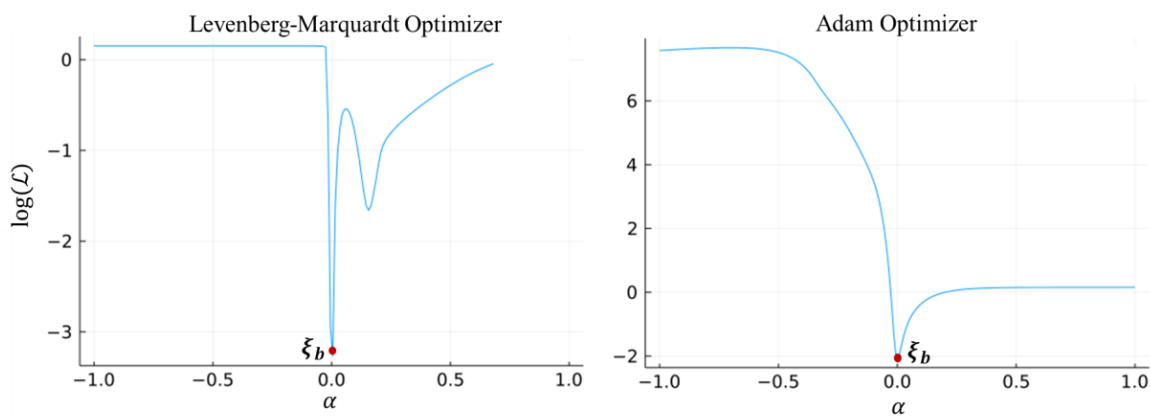

Figure 7. *Visualization of loss function surface for the optimum obtained from the LM optimizer (left) and Adam optimizer (right).*

Figure 7 shows the variation of the loss function close to the best parameters found by the LM algorithm (second-order) and Adam optimizer [43] (first-order). These plots provide a qualitative idea of the trajectories followed by the optimizers to arrive at the best parameters they attained. In the case of the LM optimizer, the surface close to $\xi_b$ is extremely non-convex with strong curvatures that sometimes look like cliffs. The LM algorithm is able to navigate these challenging topologies to arrive at a good solution, as shown in the results in section 3.1. On the other hand, Adam optimizer gets trapped in a local minimum that has a loss that is over one order of magnitude higher than that from LM. It should be noted that the optimization problem in both cases is the same. Overall, the plots demonstrate is that the LM optimizer can navigate complex topologies (characteristic of stiff chemical kinetics) to arrive at good solutions, while first-order optimizers are likely to get trapped in local minima, since they are not able to successfully navigate such complex features.

## 4. Conclusion

In this study, a novel neural ordinary differential equation approach to predict the evolution of chemical species in combustion systems was presented. The approach employed neural networks to learn the appropriate chemical source terms that lead to the correct ground-truth solution. By calculating the sensitivities of the ordinary differential equation solution to the neural network parameters, the weights and biases of the neural networks were progressively adjusted to obtain an accurate solution. The ChemNODE approach was used to learn the source terms for a zero-dimensional homogeneous constant pressure reactor with hydrogen-air combustion. The results showed that ChemNODE was able to capture the correct time evolution for all species over a wide range of conditions. It was also shown that ignition delay and its variation as a function of initial equivalence ratio and temperature was well predicted. Lastly, the results demonstrated that ChemNODE was about 2.3 times faster than the full hydrogen-air chemical mechanism, indicating its promise for providing even more significant savings if applied to higher hydrocarbon fuels with more complex chemistry and larger kinetic mechanisms. Results that demonstrate the need for optimizers that approximate or compute the curvature of the loss function surface were also presented.

As a proof-of-concept study of the novel ChemNODE approach, hydrogen-air combustion in a homogenous reactor was considered. As may be expected, ChemNODE's training process is highly dependent on the availability of efficient numerical solvers, since the integration of the ordinary differential equations is embedded in the training process. Thus, for highly stiff problems occurring in the combustion of hydrocarbons at high pressures, it is necessary to introduce stiff solvers that enable rapid integration of the governing ordinary differential equations. Further studies will include the implementation of advanced solvers within the overall framework, which will enable the extension of ChemNODE to the combustion of complex hydrocarbon fuels, and demonstration of ChemNODE in practical computational fluid dynamics simulations of gas turbine combustors and internal combustion engines via direct coupling with a computational fluid dynamics solver.

**Acknowledgments**

The submitted manuscript has been created by UChicago Argonne, LLC, Operator of Argonne National Laboratory (Argonne). The U.S. Government retains for itself, and others acting on its behalf, a paid-up nonexclusive, irrevocable world-wide license in said article to reproduce, prepare derivative works, distribute copies to the public, and perform publicly and display publicly, by or on behalf of the Government. This work was supported by the U.S. Department of Energy, Office of Science under contract DE-AC02-06CH11357. The research work was funded by Argonne's Laboratory Directed Research and Development (LDRD) Innovate project #2020-0203. The authors acknowledge the computing resources available via Blues, a high-performance computing cluster operated by the Laboratory Computing Resource Center (LCRC) at Argonne National Laboratory.